\documentstyle[preprint,prb,aps,epsfig]{revtex}
\tightenlines

\begin{document}
\draft
\begin{title}
{One-way multigrid method in electronic structure calculations}
\end{title}

\author{In-Ho Lee}
\address{School of Physics, Korea Institute for Advanced Study,
Cheongryangri-dong, Dongdaemun-gu, Seoul 130-012, Korea}

\author{Yong-Hoon Kim and Richard M. Martin}
\address{Department of Physics, University of Illinois at Urbana
Champaign, Illinois 61801}

%\date{\today}
\date{September 17, 1999}

\maketitle

\begin{abstract}
We propose a simple and efficient one-way multigrid method for
self-consistent electronic structure calculations based on iterative
diagonalization. Total energy calculations are performed on several
different levels of grids starting from the coarsest grid, with wave
functions transferred to each finer level. The only changes compared to a
single grid calculation are interpolation and orthonormalization steps
outside the original total energy calculation and required only for
transferring between grids. This feature results in a minimal amount of code
change, and enables us to employ a sophisticated interpolation method and
noninteger ratio of grid spacings.  Calculations employing a preconditioned
conjugate gradient method are presented for two examples, a quantum dot and
a charged molecular system.  Use of three grid levels with grid spacings
$2h$, $1.5h$, and $h$ decreases the computer time by about a factor of 5
compared to single level calculations.
\end{abstract}


\narrowtext

\section{Introduction}
\label{intr}

Recently, the usefulness of the real-space technique based on
three-dimensional uniform grid and higher-order finite-difference formula
\cite{FORNBERG} has been demonstrated
\cite{james,LVMJ} for the electronic structure calculations within the
framework of the Kohn-Sham (KS) density functional theory (DFT).\cite{hohen}
All the computations are performed in real space without resort to
fast Fourier transforms as in the planewave formalism.  The major parts of
calculations are local operations, so the algorithm is easily parallelized.
Furthermore, explicit storage of the Hamiltonian matrix elements can be
avoided, since the Laplacians and potential-wave functions multiplications
are respectively evaluated by the finite-difference operation on the wave
functions and a simple one-dimensional vector multiplications.

Since the number of grids $N_g$ is order of $10^5 \sim 10^6$, which
increases further with the system size and/or the level of accuracy, one
requires an efficient numerical procedure for the Hamiltonian
diagonalization in finite-difference real-space schemes.  Iterative
diagonalization methods are usually employed as in other modern electronic
structure calculations, and due to the orthogonality condition between the
KS orbitals the complexity of this iterative diagonalization scales as
$O(N_b^2 N_g)$, where $N_b$ represents the number of lowest states taken
into account. It turns out that the prefactor of this scaling can be very
dependent on the details of calculation scheme, and the development of a new
algorithm which results in the optimal prefactor is a very important and
challenging problem at the moment. Among the most promising approaches in
the literature are the multigrid (MG) algorithms.
\cite{NRF,ptvf} MG methods
originated as attempts to accelerate relaxation methods and they have been
very successful in improving the speed and accuracy in a wide variety of
science and engineering applications by combining computations at different
scales of grid levels. In the context of DFT electronic structure
calculations, several groups have already applied the multigrid scheme to
the solution of KS equations and demonstrated its efficiency: Briggs
{\em et al.} adopted coarse-grid-correction multigrid algorithm to the
calculations of various periodic and nonperiodic systems. \cite{emil}
Ancilotto {\em et al.} implemented full multigrid diagonalization procedure
to study the fragmentation of charged Li clusters. \cite{francesco}  While
these two works employed the pseudopotentials, Beck {\em et al.} has
demonstrated the feasibility of all-electron grid calculation by employing
full multigrid algorithm. \cite{Wan99}
These authors typically use integer ratio of grid spacings (e.g., $4h$, $2h$,
and $h$) and correction multilevel algorithm (V-cycle).

In this article, we introduce a simple one-way multigrid algorithm
\cite{NRF} to accelerate
self-consistent electronic structure calculations based on iterative
diagonalization. Calculations start from the coarsest grid level and
approximate solutions are transferred successively up to the finest grid. An
interesting aspect of this method is that the number of interpolation is
minimized
: Interpolations are performed outside of original total energy calculation
part, hence only for (number of grid levels $-1$) times when the wave
functions are transferred to the next finer grid level. It enables us to use
an accurate interpolation scheme and the noninteger ratio of grid spacing in
the hierarchy of grids.  Specifically we employ three different uniform
grids spacings, $2h$, $1.5h$, and $h$ to obtain the solution at the
resolution of grid spacing $h$, in which calculations on the preceding two
coarse grids provide a good initial guess of the wave functions for the
finest level calculation. We demonstrate the efficiency of the current
scheme on the twenty-electron quantum dots and the charged H cluster systems
in which the ionic potentials have been replaced by {\em ab initio}
pseudopotentials.  The comparison with a single-level calculations shows a
factor of 5 improvement in CPU time.

\section{Methods}
\label{theo}

\subsection{Basic issues}
\label{subsec:basics}

The iterative total energy minimization based on DFT is a nonlinear problem
in which the KS equations
(Hartree atomic units are used throughout this paper)
\begin{equation}
\left \{ -\frac{1}{2}\vec{\nabla}^2 + V_{ext}(\vec{r})
+ \tilde{V}_{H}(\vec{r}) + \tilde{V}_{xc}^{\sigma}(\vec{r})
\right \} \psi_j^{\sigma}(\vec{r})
= \epsilon_j^{\sigma} \psi_j^{\sigma}(\vec{r}),
\:
\sigma = \uparrow,\downarrow,
j=1,\cdots,N_b
\label{eqkohnsham}
\end{equation}
and the Poisson equation
\begin{equation}
\vec{\nabla}^2 \tilde{V}_H(\vec{r}) = -4 \pi \rho(\vec{r}),
\label{eqpoisson}
\end{equation}
are closely coupled in the self-consistency loop. \cite{ptaaj} Here
$\tilde{V}_H(\vec{r})$ and $\tilde{V}_{xc}^{\sigma}(\vec{r})$ respectively
represent the input Hartree and spin dependent exchange-correlation
potential,  at each iteration within the self-consistent calculations.
$V_{ext}(\vec{r})$ stands for the external potential, and the charge density
$\rho(\vec{r})$ is defined as the squared summation of the occupied KS
orbitals. In the higher-order finite-difference real space formulation, the
KS and Poisson equations are discretized on a uniform grid. 
The Laplacian operation is evaluated by the higher-order finite difference formula
\cite{FORNBERG} which is characterized by the finite-difference order $N$
and grid spacing $h$:
\begin{equation}
 \frac{d^2}{d x^2} f(x) = \sum_{j=-N}^{N} C_j f(x+jh) +O(h^{2N+2}),
\label{eqseco}
\end{equation}
where $\{C_j\}$ are constants.

In the present work solutions of the KS equations for the lowest
$N_b$ eigenstates are found by the
iterative preconditioned conjugate gradient method of Bylander, Kleinman,
and Lee, \cite{bkl,seit} for a given total potential  $
V_{KS}^{\sigma}(\vec{r}) = V_{ext}(\vec{r}) + \tilde{V}_{H}(\vec{r}) +
\tilde{V}_{xc}^{\sigma}(\vec{r})$. The Hartree potential $V_H(\vec{r})$ is
obtained by solving the Poisson equation. Note that, for each
self-consistent step, we need to solve two Poisson equations for the given
input and output charge densities. For the finite systems considered here
the boundary values of Hartree potential are evaluated using a multipole
expansion of the potential of the charge distribution and the relaxation
vectors at the boundary are set to zero for the Dirichlet boundary
conditions. The solution of Poisson equation
inside of the box has been first generated by a Fourier method with low
order finite difference ($N=1$), \cite{ptvf} and it has been subsequently
relaxed by the preconditioned conjugate gradient method \cite{john,hoshi}
with higher-order finite difference formula.  At each step we choose the new
input density and potential using a simple linear mixing of output and input
densities. \cite{bkl}

After obtaining orbitals and density from self-consistent solutions of
Poisson and KS equations, the total electronic energy is obtained:
\begin{eqnarray}
E_{tot} &=& \sum_{\sigma,j} \epsilon_{j}^{\sigma}
-\sum_{\sigma} \int d^3r \{ \tilde{V}_H(\vec{r})
+\tilde{V}_{xc}^{\sigma}(\vec{r})  \} \rho^{\sigma}(\vec{r})
+\frac{1}{2}  \int \! d^3r' \! \int \! d^3r \frac{\rho(\vec{r}) \rho(\vec{r}') }
{\mid \vec{r}-\vec{r}' \mid}
+E_{xc}[\rho^{\uparrow}(\vec{r}),\rho^{\downarrow}(\vec{r})],
\end{eqnarray}
where the summations over the single particle energy
($\epsilon_{j}^{\sigma}$) are carried out for all the states below the Fermi
level, and  $E_{xc}[\rho^{\uparrow}(\vec{r}),\rho^{\downarrow}(\vec{r})]$ is
the exchange-correlation energy.  For our local spin density approximation
we use the Perdew and Zunger's parameterization of the Ceperley and
Alder's quantum Monte Carlo data. \cite{ca}

\subsection{One-way multigrid method}
\label{subsec:OWMG}

The most time-consuming part of the self-consistent electronic structure
calculations described in the previous subsection is the iterative solution
of KS equations.
The sources of this computation bottleneck can be traced to broadly two (but
closely related) aspects of self-consistent iterative diagonalization
schemes.  First of all, in general we do not have a good initial guess of wave
functions, which generate density, and hence $\tilde{V}_H(\vec{r})$ and
$\tilde{V}_{xc}^{\sigma}(\vec{r})$ in Eq. ({\ref{eqkohnsham}).   So initial
several self-consistency steps will be used to obtain solutions of biased
Hamiltonians, although they tend to be the most time-consuming part.
Secondly, in single iterative solution of KS equations, a direct
application of a relaxation method on the fine grid has trouble in damping
out the long-ranged or slowly varying error components in the orbitals. This
can be understood by the usual spectral analysis of relaxation scheme, or
considering that the nonlocal Laplacian operation on a fine grid is
physically short-ranged.  This means that there is an imbalance in the
relaxation step for the long-ranged and short-ranged error components.

MG is a quite general concept, and the choice of a specific algorithm is
very dependent on the problem under consideration. For our purpose, we seek
a procedure
which generates a good initial guess for the finest grid calculation and
effectively removes long-range error components of wave functions in the
solution of KS equations.
In this work, we employed the one-way multigrid scheme with three different
uniform grids with noninteger ratio of spacings, $2h$, $1.5h$, and $h$.
The calculation starts from the coarsest grid $2h$, and in each grid-level
calculation,  Eqs. (\ref{eqpoisson}) and (\ref{eqkohnsham}) are solved
self-consistently as in the usual single-level algorithm.  After each
self-consistent calculation on a coarse grid, only wave functions are
interpolated to the next fine grid, and another set of self-consistent
calculation is performed.
Since that the interpolated wave functions usually do not satisfy the
orthonormality condition any more, we take an extra Gram-Schmidt
orthogonalization process after each orbital interpolation. So we have two
interpolations and two Gram-Schmidt orthogonalization processes for our
hierarchy of three grid systems. In Fig. \ref{owmg}, we summarize the
algorithmic flow of the procedures.

While an efficient interpolation/projection scheme is a crucial ingredient
of any successful application of MG method, we note that it can be also
time-consuming and tricky part due to the physical conditions such as
orthonormality of wave functions. Hence our strategy, which is the
characteristic of the current scheme, is to minimize the number of data
transfer between different grid levels, while employ a sophisticated
interpolation method which is very accurate and allow us to use a noninteger
ratio of grid spacings. Specifically, we used a three-dimensional piecewise
polynomial interpolation with a tensor product of one-dimensional B-splines
as the interpolating function.
\cite{ptvf,deboor} A piecewise cubic polynomials have been taken as B-splines.

\section{Efficiency and discussions}
\label{resdis}

We consider two different electronic systems of a localized quantum dot
model and a charged hydrogen cluster to demonstrate the efficiency of the
present algorithm.  We first take a twenty-electron quasi two-dimensional
quantum dot modeled by an anisotropic parabolic confinement
potential\cite{LVMJ} $V_{ext}(\vec{r})= \frac{1}{2} \omega_x^2 x^2
+\frac{1}{2} \omega_y^2 y^2 +\frac{1}{2} \omega_z^2 z^2$.  In-plane
potential is characterized by the confinement energies $\omega_x=\omega_y=5$
meV, and $\omega_z=45$ meV has been taken to reproduce the dot-growth
$z$-direction confinement caused by the quantum wells and heterojunctions.
\cite{LVMJ} Our calculations for anisotropic parabolic dot in GaAs host
material (dielectric constant $\epsilon=12.9$, effective mass $m^*=
0.067m_e$) are based on the effective mass approximation, and rescaled
length and energy units are $a_B^*=101.88$ {\AA} and 10.96 meV,
respectively.
Uniform grid spacing $h=0.3a_B^*$ with box size $ 81 \times 81 \times 21
a_B^{* \ 3}$ have been used, hence the number of grid points is about $1.4
\times 10^5$ points at the finest grid level ($h = 0.3 a_B^*$) while only
about $1.6 \times 10^4$ points at the coarsest grid level ($2h = 0.6
a_B^*$).  Finite difference order $N=3$ has been used at grid levels $h$ and
$1.5h$, while $N=1$ for grid level $2h$, to solve a set of spin-polarized
KS equations with fifteen orbitals in each spin channel. Noninteracting
eigenstates (Hermite polynomials) are used as an initial guess for the
coarsest grid calculation.

The CPU times for each self-consistent iteration are shown in Fig.
\ref{cpu1}. The horizontal axis stands for the self-consistency iteration
index, while the vertical axis is the required computer time for a given
iteration step.  The case of the present three-level one-way multigrid
algorithm is shown in the lowest panel (c).  Comparing with the results of a
single-level calculation shown in the panel (a), we see
significant savings of the computation time, in which total computation time
is about 5 times shorter than a single-level calculation.  While the
three-level MG scheme requires more number of self-consistent iterations (28
iterations compared with 20 iterations), they are mostly performed in the
coarsest grid $2h$, and at the finest grid level $h$ we only need several
iterations. Interpolation and orthonormalization steps are indicated by
downward arrows, which take only a small amount of computation time.

To further demonstrate the advantage of the usage of the intermediate grid
spacing $1.5h$ in our three-level scheme, we show the CPU time of two-level
($2h$ and $h$) calculation in the panel (b).  While the number of iterations
taken in the finest grid $h$ is still much smaller than the single-level
calculation, it is much larger than that in the three-level calculation,
resulting in the ratio of computation times $5:2:1$ for one-, two-, and
three-level grid calculations.  Although noninteger ratio of grid spacing is
not widely used in MG applications, this clearly shows its usefulness.

We obtained similar results of factor 5 improvements in computation
speed for other test cases, such as the {\em ab initio} nonlocal
pseudopotential calculation of a charged hydrogen cluster H$_9^+$.  Details
of the calculation are identical to those of quantum dot calculations,
except that ionic external potentials are treated by separable \cite{kb}
nonlocal pseudopotential generated by the method of Troullier and Martins.
\cite{tm}
Finite difference order $N=6$ at grid $h$ and $1.5h$, and $N=1$ for the
grid $2h$ have been used to solve spin-unpolarized KS equations with the
lowest 6 states. The number of grid points involved in the finest grid
calculation is $3.5 \times 10^5$, while it is $4.3 \times 10^4$ for the
coarsest grid calculation.

We have to emphasize that the improvements seen in previous examples are
purely induced by the a simple usage of MG idea, in which the only
modification from the original single-level code was the addition of an
outer loop which transfer the wave functions.  We can expect that the
introduction of the MG scheme at different stages of calculations, such as
the correction path for the relaxation of KS orbital or Hartree potential,
will result in further improvements.
To do so, we will need additional residual computation and projection steps
that can be combined with our conjugate gradient solvers.
We also note that it will require an interpolation strategy and grid levels
which are different from the current method.  Finally, we point out that
this type of one-way multigrid idea is very similar to often-used practices
in plane-wave calculations based on iterative diagonalization, in which a
solution is first found at one energy cutoff (equivalent to a coarse grid)
and used as the input to a higher energy cutoff calculation (equivalent to a
finer grid).
This corresponds to interpolating solutions from a coarser to a finer grid
using Fourier components.

\section{Conclusions}
\label{cnclsn}

In this work, we demonstrated that the introduction of a simple one-way
multigrid method greatly improves the efficiency of real-space electronic
structure calculations based on the iterative solution of KS equations.
While minimizing the number of data transfer between grids, we employed
an accurate interpolation method, which enabled us to incorporate
three-level grids with noninteger ratio of grid spacings.  Our general
strategy of using $2h$, $1.5h$, and $h$, showed a factor 5 improvement of
computation time, while it required only minimal computer code
modifications.
The usefulness of the intermediate grid step $1.5h$ has been shown by
comparing the current scheme with two-level ($2h$ and $h$) calculations.

\acknowledgments
This work was supported in part by the National
Science Foundation under grant DMR 9802373.
We are grateful to supercomputer center SERI.

\begin{figure}
\caption{ Schematic diagram of the present one-way multigrid algorithm
discussed in the text. The self-consistent calculation at each level is done
by using preconditioned conjugate gradient relaxation. The values in
circles, $2h$, $1.5h$, and $h$ stand for the uniform grid spacing for a
given level. The calculation starts at the coarsest level (level 1, $2h$) at
the bottom, and ends at the finest grid (level 3, $h$) at the top. Orbital
interpolation and orthogonalization step is taken after each coarse grid
(level 1 and 2) calculation. }
\label{owmg}
\end{figure}

\begin{figure}
\caption{ CPU time vs. self-consistent iteration number for 20-electron
quantum dot calculations in (a) single-level ($h$), (b) two-level ($2h$ and
$h$), and (c) three-level ($2h$, $1.5h$, and $h$) schemes. Within the local
spin density approximation, we minimized the total energy with respect to
the electronic degree of freedom.
Downward arrows in (b) and (c) indicate the interpolation-orthogonalization
steps. Calculations are performed on a DEC alpha 433au personal workstation.
}
\label{cpu1}
\end{figure}

\newpage
\begin{minipage}[H]{0.60\linewidth}
\vspace{1.0cm}
\centering\epsfig{file=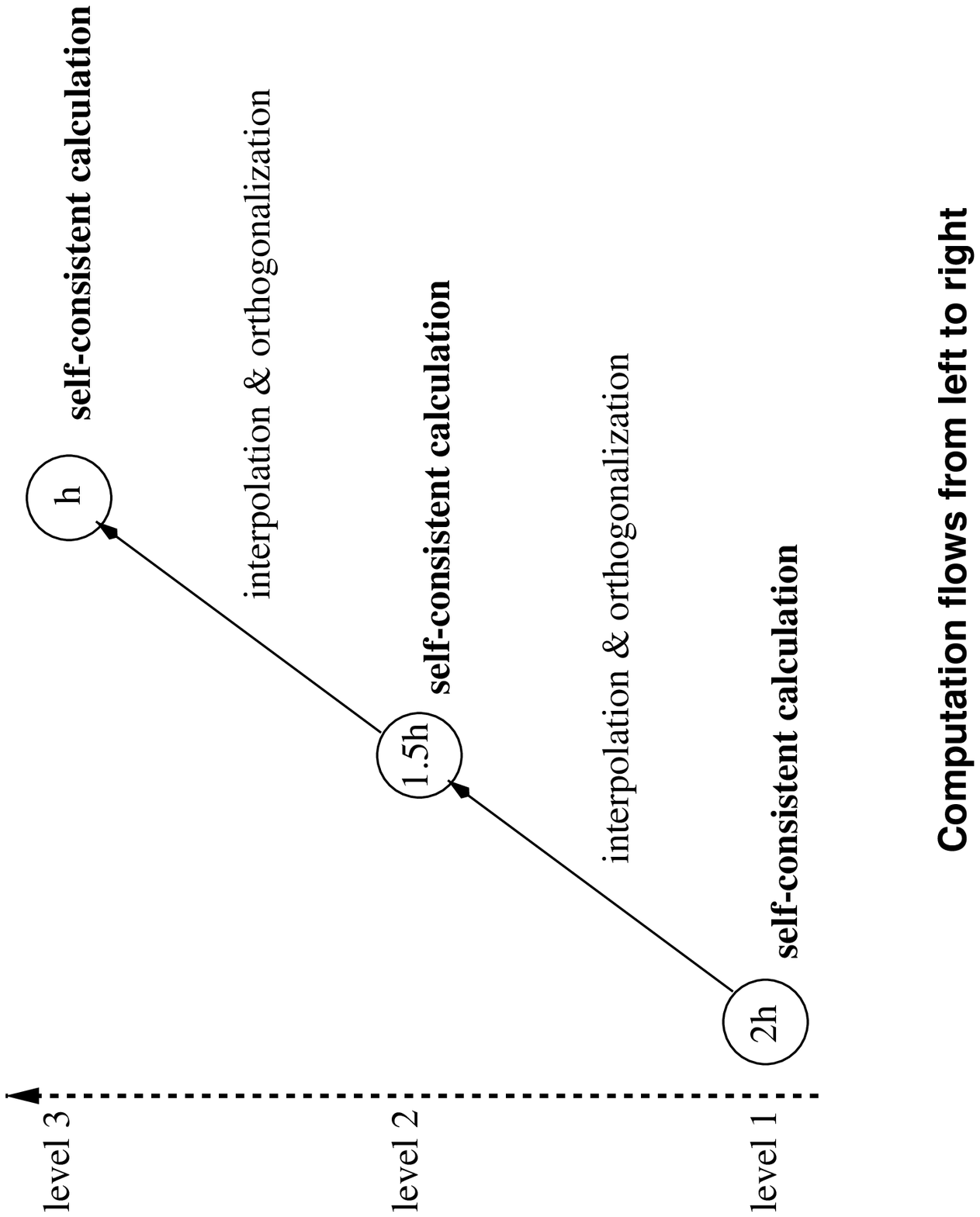,width=\linewidth}
\end{minipage} \hspace{0.2cm}
Fig. \ref{owmg}

\newpage
\begin{minipage}[H]{0.60\linewidth}
\vspace{1.0cm}
\centering\epsfig{file=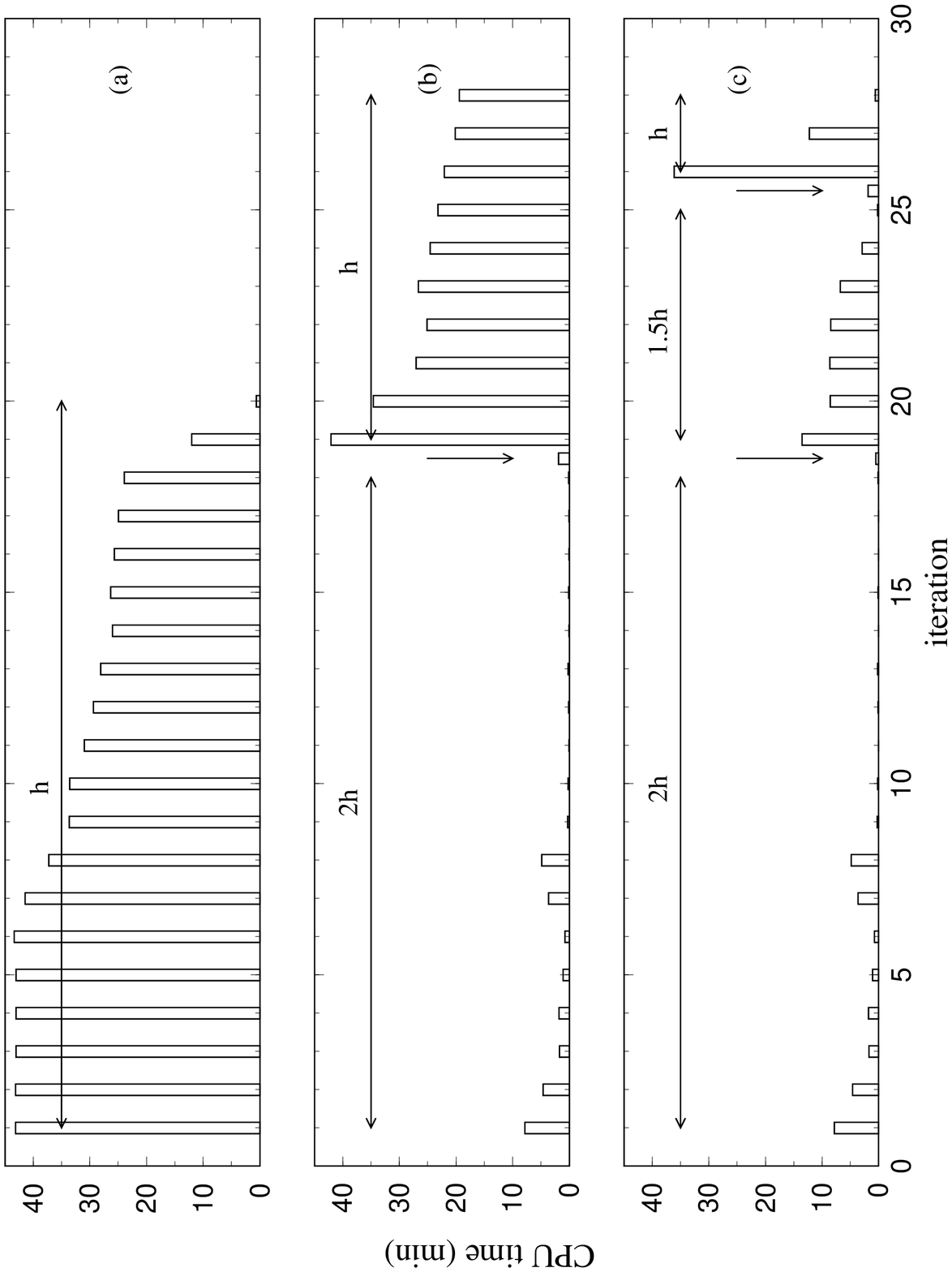,width=\linewidth}
\end{minipage} \hspace{0.2cm}
Fig. \ref{cpu1}

\end{document}